\documentclass[aps,prd,onecolumn,showpacs,superscriptaddress,showkeys,nofootinbib]{revtex4}
\usepackage{amsfonts}

\usepackage{amsmath}
\usepackage{graphicx}
\usepackage[latin9]{inputenc}
\usepackage{amsmath}
\usepackage{amssymb}

\newcommand{\bastar}{\begin{eqnarray*}}
\newcommand{\eastar}{\end{eqnarray*}}
\newskip\humongous \humongous=0pt plus 1000pt minus 1000pt

\newif\ifdtup

\relax

\begin{document}

\title{Gravitational catalysis of merons in Einstein-Yang-Mills theory}
\author{Fabrizio Canfora}
\email{canfora@cecs.cl}
\affiliation{Centro de Estudios Cientificos (CECS), Casilla 1469, Valdivia, Chile}
\author{Seung Hun Oh}
\email{shoh.physics@gmail.com}
\affiliation{Department of Physics, Konkuk University, Seoul 05029, Korea}
\author{Patricio Salgado-Rebolledo}
\email{patricio.salgado@uai.cl}
\affiliation{Facultad de Ingeniería y Ciencias \& UAI Physics Center, Universidad Adolfo
Ibanez, Avda. Diagonal Las Torres 2640, Peñalolen, Santiago, Chile}

\begin{abstract}
We construct regular configurations of the Einstein-Yang-Mills theory in
various dimensions. The gauge field is of meron-type: it is proportional to
a pure gauge (with a suitable parameter $\lambda$ determined by the field
equations). The corresponding smooth gauge transformation cannot be deformed
continuously to the identity. In the three-dimensional case we consider the
inclusion of a Chern-Simons term into the analysis, allowing $\lambda$ to be
different from its usual value of $1/2$. In four dimensions, the gravitating
meron is a smooth Euclidean wormhole interpolating between different vacua
of the theory. In five and higher dimensions smooth meron-like
configurations can also be constructed by considering warped products of the
three-sphere and lower-dimensional Einstein manifolds. In all cases merons
(which on flat spaces would be singular) become regular due to the coupling
with general relativity. This effect is named ``gravitational catalysis of
merons''.
\end{abstract}

\maketitle

\bigskip

\section{Introduction}

The existence of topological solitons is one of the most important
non-perturbative effects in field theory \cite{manton}. These non-trivial
topological objects are believed to play a fundamental role in the color
confinement problem (for a detailed review, see \cite{confinement}) which is
one of the \textquotedblleft big\textquotedblright\ open issues in gauge
field theory. A very important class of topological solitons is the
Euclidean one (namely, regular solutions of the Euclidean theory). Euclidean
topological solitons are especially relevant as they play a very important
role at quantum level as non-trivial saddle points of the path integral. The
most important Euclidean solutions are instantons (which are local regular
minima of the Euclidean action) and sphalerons (which are saddle points with
one-or a finite number of-unstable mode(s)). Unfortunately, analytic
solutions are available only in special cases (in particular, when suitable
BPS bounds can be saturated). In the case of instantons of Yang-Mills theory
in 4 dimensions the saturation of the bound is equivalent to the
self-duality condition. From the point of view of gravitational
back-reaction, instantons are not very interesting as the self-duality
condition implies that the energy-momentum tensor of the self-dual instanton
vanishes so that it does not back-reacts on the metric at semi-classical
level. From the Yang-Mills point of view, a very important type of Euclidean
configurations are the so-called merons, firstly introduced in \cite{meronp1}%
. Merons are gauge fields interpolating between different topological sectors%
\footnote{%
One of the results of the present paper, as it will be explained in the next
sections, is to construct a quite remarkable and concrete confirmation of 
this interpretation in the gravitating case.}. In particular, instantons can
be interpreted as merons bound states \cite{meronp2,meronp3,meronp4,meronsp5}%
. It is commonly accepted that merons are quite relevant configurations from
the point of view of the confinement problem (see, for instance, \cite%
{confinement} \cite{meronp4}). In flat Euclidean spaces, merons are usually
singular. Hence, on flat Euclidean spaces, a single \textquotedblleft
isolated\textquotedblright meron gives a vanishing contribution to the path
integral as its Euclidean action is divergent. It is well known that merons
are relevant only as \textquotedblleft building blocks\textquotedblright\ of
the instantons in the usual cases.

It is quite obvious that in many physically relevant situations the coupling
with Einstein gravity\footnote{
Or, at the very least, the non-vanishing curvature of space-time.} cannot be
neglected (this is the case for instance in early cosmology \cite{rubakov}
when topological solitons are believed to play a fundamental role).
Consequently, a very important question arises: Is it still true that merons
are necessarily singular even when the coupling with General Relativity is
taken into account? Indeed, due to the reasons mentioned above, whether or
not merons are singular\footnote{
Hence, whether or not merons can give a finite contribution to the 
semi-classical path integral through the corresponding saddle points.} can
have a big influence on our understanding of the confinement problem. A
first hint that the coupling of merons with general relativity can change
the \textquotedblleft flat\textquotedblright\ picture quite considerably can
be found (with Lorentzian signature) in \cite{teitelboim} \cite{mbh} where
it has been shown that the singularity of the simplest meron can be hidden
behind a black hole horizon.

A further very important situation where topological solitons play a
fundamental role is in three Euclidean dimensions. The interest of the
3-dimensional case lies in the fact that difficult non-perturbative
questions are easier to understand in three-dimensional Yang--Mills theory
than in the four dimensions. Despite being simpler than QCD,
three-dimensional Yang-Mills theory possesses local interacting degrees of
freedom. A further benefit of three-dimensional Yang-Mills theory is that it
is a good approximation of high temperature QCD\footnote{
In which case the mass gap plays the role of the magnetic mass}. Last but
not least, the Chern--Simons term can be included \cite{tmym1,tmym2},
leading to a mass for the gauge field which is of topological origin. The
inclusion of the Chern-Simons term is not only a nice theoretical exercise
since it can be shown that such a term appears upon integrating out the
fermions (see, for instance \cite{CSF1}\ and \cite{CSF2}; a detailed review
is \cite{reviewCS}). Moreover, the non-perturbative features of
topologically massive Yang-Mills theory in three dimensions are in a very
good agreement with the expected confinement picture \cite{ourann}.

Very deep open issues related to three-dimensional topologically massive
Yang-Mills theory are related to the following fact. Such a theory in a
suitable range of parameters (see \cite{ourann}) is confining. Standard
arguments (see \cite{confinement}) suggest that regular non-trivial
Euclidean saddle points of the path integral must play a fundamental role to
understand confinement. However, in three Euclidean dimensions, it is not
possible to construct the usual self-dual Yang-Mills instantons (since one
would need the four-dimensional Levi-Civita $\varepsilon$-symbol). In fact,
as it will be discussed in the next sections, although there are no
self-dual instantons in three Euclidean dimensions one can still construct
regular smooth gravitating merons.

In general, it is very difficult to analyze the gravitational properties of
topologically non-trivial configurations. Due to the difficulties in
constructing analytic regular configurations of the four-dimensional
Einstein-Yang-Mills system many of the available results are numerical (see,
for instance, \cite{numym1,numym2,numym3,numym4,numym5}).

The first aim of the present paper is to show that, nevertheless, it is
possible to construct analytic regular solutions corresponding to
gravitating merons in various dimensions in Euclidean Einstein-Yang-Mills
theory. In order to achieve this goal two techniques are combined. The first
technique is based on the $SU(2)$-valued generalized hedgehog ansatz
(introduced in \cite%
{canfora,canfora2,canfora3,canfora4,canfora4bis,yang1,canfora5,yang2,canfora6,canfora7,canfora8,canfora9,canfora10,canfora11,canfora12,canfora13}%
), which works both for the Skyrme model and for the Yang-Mills-Higgs
system. The second is based on the Cho approach \cite%
{cho1,cho2,cho3,cho4,cho5}.

The second aim is to show the coupling with Einstein gravity can change
quite dramatically the usual physical interpretation of merons. In the
three-dimensional case, we construct regular gravitating meron-like
configurations and include a Chern-Simons term into the analysis. Due to the
fact that in three dimensions it is not possible to define self-dual
configurations, the regular Euclidean saddle points constructed here are
likely to play a fundamental role to understand the non-perturbative
features of the theory. In the four-dimensional case, we construct different
regular gravitating meron-like configurations. Such configurations can be
seen as smooth Euclidean wormholes interpolating between different vacua of
the theory. Euclidean wormholes \cite{maldacena, hosoya, gupta, rey,
donets,euclidw1,euclidw2,euclidw3,euclidw4,euclidw5,euclidw6,euclidw7,euclidw8}
(see, for a recent view on this topic, \cite{euclidw9}) can be defined as
extrema of the action in Euclidean quantum gravity connecting distant
regions. It is widely recognized that such configurations can have quite
remarkable physical consequences (as discussed in details in the above
references). In five dimensions dimensions we construct regular meron-like
configurations that generalize the three-dimensional result previously found
for $\lambda =1/2$. The metric is given by the a two-dimensional constant
curvature space times the three-sphere. This result can be further extended
to arbitrary higher dimensions. In dimension $D>6$ the metric turns out to
be given by the warped product of the three-sphere and any solution of the $%
D-3$-dimensional Einstein equations in vacuum with an effective cosmological
constant.

This paper is organized as follows: in the second section, meron-like
configurations within the Euclidean Einstein-Yang-Mills theory are
introduced. In the third section, we present a general ansatz to construct
merons and Einstein-Yang-Mills equations are discussed. In the fourth
section the solutions are constructed. First, three-dimensional smooth
regular gravitating merons are considered, the effects of the Chern-Simons
term are included and the corresponding Euclidean action is computed. In
four-dimensional case, smooth and regular gravitating merons are presented
and their interpretation as Euclidean wormholes is discussed. Finally, we
construct regular meron-like configurations in five and higher dimensions.
In the fifth section, some conclusions are drawn.

\section{The System}

We consider the Euclidean Einstein-Yang-Mills system in $D$ dimensions with
cosmological constant. The action of the system is 
\begin{equation}
S=S_{\mathrm{G}}+S_{\mathrm{SU(2)}}\ ,  \label{skyoo}
\end{equation}
where the gravitational action $S_{\mathrm{G}}$ and the gauge field action $%
S_{\mathrm{SU(2)}}$ are given by 
\begin{align}
S_{\mathrm{G}}= & -\frac{1}{16\pi G}\int d^{D}x\sqrt{g}(R-2\Lambda)\,,
\label{sky1o} \\
S_{\mathrm{SU(2)}}= & -\frac{1}{8e^{2}}\int d^{D}x\sqrt{g}\mathrm{Tr}%
\left(F^{\mu\nu}F_{\mu\nu}\right)\,.  \label{sky2o}
\end{align}
where R is the Ricci scalar, $G$ is Newton's constant, $\Lambda$ is the
cosmological constant, $F_{\mu\nu}
=\partial_{\mu}A_{\nu}-\partial_{\nu}A_{\mu}+\left[A_{\mu},A_{\nu}\right]$
is the field strength associated to the gauge field $A_{\mu}$ and $e$ is the
Yang-Mills coupling constant. In our conventions $c=\hbar=1$. The resulting $%
N$-dimensional Einstein equations are 
\begin{equation}
G_{\mu\nu}+\Lambda g_{\mu\nu}=8\pi GT_{\mu\nu}\,,  \label{einstein}
\end{equation}
where $G_{\mu\nu}$ is the Einstein tensor and $T_{\mu\nu}$ is the
stress-energy tensor of the Yang-Mills field 
\begin{equation}
T_{\mu\nu}=\frac{2}{\sqrt{g}}\frac{\delta S_{\mathrm{SU(2)}}}{\delta
g^{\mu\nu}}=-\frac{1}{2e^{2}}\mathrm{Tr}\left(F_{\mu\alpha}F_{\nu\beta}g^{
\alpha\beta}-\frac{1}{4}g_{\mu\nu}F^{\rho\sigma}F_{\rho\sigma}\right)\,.
\label{T1}
\end{equation}
The Yang-Mills equations are given by 
\begin{equation}
\mathrm{YM}^{\mu}=\nabla_{\nu}F^{\mu\nu}+\left[A_{\nu},F^{\mu\nu}\right]=0\ ,
\label{YM1}
\end{equation}
where $\nabla^{\mu}$ is the Levi-Civita covariant derivative. The connection 
$A_{\mu}=A_{\mu}^{A}t_{A}$ takes values on the $SU(2)$ algebra, whose
generators are defined as 
\begin{equation}
t_{A}=i\sigma_{A}\,\, ,\,\,A=1,2,3\,,\,
\end{equation}
$\sigma_{A}$ being the Pauli matrices.

Meron-like configurations as well as their important role in the
non-perturbative sector of Yang-Mills theory have been extensively discussed
in the literature (see, for instance, \cite%
{meronp1,meronp2,meronp3,meronp4,meronsp5}). All the most important examples
can be written in the following form\footnote{
It is more common to use the 't Hooft symbol (which is a Levi-Civita $
\varepsilon $-tensor in which some of the indices are internal while other 
are space-time indices). On flat spaces, the usual notation is equivalent to
the one in Eq. (\ref{merona1}). On curved spaces the notation in Eq. (\ref%
{merona1}) is much more convenient as it avoids the problem to properly 
define the 't Hooft symbol on curved spaces.}\cite{meronsp5} 
\begin{equation}
A_{\mu }=\lambda U^{-1}\partial _{\mu }U\ ,\ \lambda \neq 0,1\ .
\label{merona1}
\end{equation}%
As it will be shown in the following, the Yang-Mills equations fix the
parameter $\lambda $. Therefore, our definition of meron in the present
paper will be a regular configuration of the form in Eq. (\ref{merona1})
constructed with a topologically non-trivial $SU(2)$ map $U(x^{\mu })$. Note
that the definition of meron in Eq. (\ref{merona1}) works both with
Euclidean and with Lorentzian signature. Although we will focus in this work
mainly on the Euclidean case, many of the present results can be easily
extended to the Lorentzian case.

We adopt the standard parametrization of the $SU(2)$-valued scalar $%
U(x^{\mu})$ 
\begin{equation}
U^{\pm1}(x^{\mu})=Y^{0}(x^{\mu})\mathbb{\mathbf{I}}\pm Y^{A}(x^{\mu})t_{A}\
,\ \ \left(Y^{0}\right)^{2}+Y^{A}Y_{A}=1\,,  \label{standnorm}
\end{equation}
where $\mathbb{\mathbf{I}}$ is the $2\times2$ identity. The last equality
implies that $(Y^{0},Y^{A})$ is a unit vector in a three sphere, which is
naturally accounted for by writing 
\begin{equation}  \label{pions}
\begin{array}{lcl}
Y^{0} =\cos\alpha\ ,\,\,\,Y^{A}=n^{A}\cdot\sin\alpha\,, &  &  \\[5pt] 
n^{1} =\sin\Theta\cos\Phi\ ,\, n^{2}=\sin\Theta\sin\Phi\ ,\,
n^{3}=\cos\Theta\ . &  & 
\end{array}%
\end{equation}
As it will be explained in the next sections, the ansatz for the $\alpha$, $%
\Theta$ and $\Phi$ functions will be chosen in order to have a non-vanishing
winding number.

\section{Ansatz}

For our purposes it will be convenient to introduce the left-invariant
Maurer-Cartan forms on $SU(2)$, which can be defined in terms of the Euler
angles $x^{i}=(\psi,\theta,\varphi)$ by 
\begin{eqnarray*}
\Gamma_{1} & = & \frac{1}{2}\left(\mathrm{sin}\psi d\theta-\mathrm{sin}%
\theta \mathrm{cos}\psi d\varphi\right), \\
\Gamma_{2} & = & \frac{1}{2}\left(\mathrm{-cos}\psi d\theta-\mathrm{sin}
\theta\mathrm{sin}\psi d\varphi\right), \\
\Gamma_{3} & = & \frac{1}{2}\left(d\psi+cos\theta d\varphi\right),
\end{eqnarray*}
\begin{eqnarray*}
0\leq\psi<4\pi\,, & \,\,0\leq\theta<\pi\,, & \,\,0\leq\varphi<2\pi\,.
\end{eqnarray*}
We will consider a $D$-dimensional euclidean space-time of the form 
\begin{equation}
ds^{2}=g_{\mu\nu}dx^{\mu}dx^{\nu}=\gamma_{ab}(z)dz^{a}dz^{b}+\rho(z)^2%
\sum_{i=1}^{3}\Gamma_{i}\otimes\Gamma_{i}\,,  \label{eq:Metric}
\end{equation}
where we have split the coordinates as $x^{\mu}=(z^{a},x^{i})$, $%
a=1,\ldots,d=D-3$, $\gamma_{ab}$ is a $d$- dimensional metric and $\rho(z)$
is a warping factor depending on the coordinates $z^{a}$ only.

As it has been discussed in \cite{canfora6}, \cite{canfora12}, \cite%
{canfora13}, the following choice for the functions in (\ref{pions}) is
suitable for the class of metrics (\ref{eq:Metric}): 
\begin{equation}
\Phi =\frac{\psi +\varphi }{2}\ ,\ \ \tan \Theta =\frac{\cot \left( \frac{
\theta }{2}\right) }{\cos \left( \frac{\psi -\varphi }{2}\right) }\ ,\ \
\tan \alpha =\frac{\sqrt{1+\tan ^{2}\Theta }}{\tan \left( \frac{\psi
-\varphi }{2}\right) }\ .  \label{pions2.25}
\end{equation}%
It is easy to verify directly that\textit{\ in any background metric of the
form in Eq}. (\ref{eq:Metric}), a meron ansatz of the form in Eqs. (\ref%
{merona1}), (\ref{pions}) and (\ref{pions2.25})\textit{\ identically
satisfies the Lorentz gauge condition} (something which simplifies
considerably the Yang-Mills equation): 
\begin{equation}
\nabla ^{\mu }A_{\mu }=0\ .
\end{equation}%
It is also worth to emphasize that the present ansatz is topologically
non-trivial as it has a non-trivial winding number along the $z^{a}=\mathrm{%
\ const}$ hypersurfaces of the metric in Eq. (\ref{eq:Metric}): 
\begin{equation}
W=-\frac{1}{24\pi ^{2}}\int_{S^{3}}tr\left[ \left( U^{-1}dU\right) ^{3}%
\right] =-\frac{1}{2\pi ^{2}}\int \sin ^{2}\alpha \sin \Theta d\alpha
d\Theta d\Phi =1\ .
\end{equation}%
Hence, the present configuration cannot be deformed continuously to the
trivial vacuum.

\subsection{Yang-Mills equations}

In the coordinates $x^{\mu}=(z^{a},x^{i})$, the gauge potential is split in
two parts $A_{\mu}=\{A_{a},A_{i}\}$. The ansatz in Eqs. (\ref{merona1}), (%
\ref{pions}) and (\ref{pions2.25}) leads to the following form for $A_{i}$ 
\begin{equation}  \label{Ameron}
\begin{array}{lcl}
A_{\psi} & = & -\dfrac{\lambda}{2}\big(\sin\theta\cos\varphi\
t_{1}+\sin\theta\sin\varphi\ t_{2}-\cos\theta\ t_{3}\big)\,, \\[5pt] 
A_{\theta} & = & \dfrac{\lambda}{2}\big(\sin\varphi\ t_{1}-\cos\varphi\ t_{2}%
\big)\,, \\[5pt] 
A_{\varphi} & = & \dfrac{\lambda}{2}\ t_{3}\,,%
\end{array}%
\end{equation}
while the components $A_{a}$ identically vanish 
\begin{equation*}
A_{a}=0\,.
\end{equation*}
As the connection is time independent, the non-Abelian ``electric'' field
vanishes and this meron-like configuration is purely ``magnetic''. In fact,
the non-vanishing space-time components of the field strength are 
\begin{equation}  \label{Fmeron}
\begin{array}{lcl}
F_{\psi\theta} & = & -\dfrac{\lambda(\lambda-1)}{2}\big(\cos\theta\cos%
\varphi\ t_{1}+\cos\theta\sin\varphi\ t_{2}+\sin\theta\ t_{3}\big)\,, \\%
[5pt] 
F_{\psi\varphi} & = & \dfrac{\lambda(\lambda-1)}{2}\sin\theta\big(%
\sin\varphi\ t_{1}-\cos\varphi\ t_{2}\big)\, , \\[5pt] 
F_{\theta\varphi} & = & \dfrac{\lambda(\lambda-1)}{2}\big(\cos\varphi\
t_{1}+\sin\varphi\ t_{2}\big)\,,%
\end{array}%
\end{equation}
and the left hand sides of Yang-Mills equations (\ref{YM1}) become, 
\begin{equation}  \label{eq:ymeqs}
\begin{array}{lcl}
\mathrm{YM}^{\psi} & = & \dfrac{8\lambda\left(\lambda-1\right)}{\rho^{4} 
\mathrm{sin}\theta}\left(2\lambda-1\right)\left(\mathrm{cos}\varphi t_{1}+%
\mathrm{sin}\varphi t_{2}\right)\,, \\[5pt] 
\mathrm{YM}^{\theta} & = & \dfrac{8\lambda\left(\lambda-1\right)\left(2
\lambda-1\right)}{\rho^{4}}\big(-\mathrm{sin}\varphi t_{1}+\mathrm{cos}%
\varphi t_{2}\big)\,, \\[5pt] 
\mathrm{YM}^{\varphi} & = & -\dfrac{8\lambda(\lambda-1)\left(2\lambda-1%
\right)}{ \rho^{4}\mathrm{sin}\theta}\big(\mathrm{cos}\theta\mathrm{cos}%
\varphi\ t_{1}+\mathrm{cos}\theta\mathrm{sin}\varphi\ t_{2}+\mathrm{sin}%
\theta\ t_{3}\big)\,, \\[5pt] 
\mathrm{YM}^{a} & = & 0\,.%
\end{array}%
\end{equation}
Therefore, the Yang-Mills equations are identically satisfied for 
\begin{equation}
\lambda=\frac{1}{2}\,.  \label{eq:solym}
\end{equation}
This is the standard value of $\lambda$ for meronic configurations (\ref%
{merona1}). As we will show, in three-dimensions it is possible to find a
different result for $\lambda$ when a Chern-Simons term is included in the
action for the SU(2) gauge field. For $D>3$ however, $\lambda=\frac{1}{2}$
will be assumed.

\subsection{Einstein equations}

In (\ref{eq:Metric}), the metric $g_{\mu\nu}$ splits as $g_{ij}=\rho(z)^{2}
h_{ij}(x)$ , $g_{ab}=\gamma_{ab}(z)$ , $g_{ia}=0$, where $h_{ij}$ is the
metric of the three sphere in the coordinates $x^{i}$, 
\begin{equation*}
\sum_{i=1}^{3}\Gamma_{i}\otimes\Gamma_{i}=h_{ij}dx^{i}dx^{j}=\frac{1}{4}%
\left(d\psi^{2}+2\cos\theta d\psi d\varphi+d\theta^{2}+d\varphi^{2}\right).
\end{equation*}
The Ricci tensor and the Ricci scalar are then given by 
\begin{eqnarray}
R_{ij} & = & 2h_{ij}\left(1-\tilde{\nabla}_{a}\rho\tilde{\nabla}^{a}\rho-%
\frac{1}{2}\rho\tilde{\nabla}^{2}\rho\right)\,,  \notag \\
R_{ia} & = & 0\,,  \notag \\
R_{ab} & = & \tilde{R}_{ab}-\frac{3}{\rho}\tilde{\nabla}_{b}\tilde{\nabla}%
_{a}\rho\,,  \label{eq:ricci} \\
R & = & \tilde{R}+\frac{6}{\rho^{2}}\left(1-\tilde{\nabla}_{a}\rho\tilde{
\nabla}^{a}\rho-\rho\tilde{\nabla}^{2}\rho\right)\,.  \notag
\end{eqnarray}
where $\tilde{R}{}_{ab}$, $\tilde{R}$ and $\tilde{\nabla}$ denote the Ricci
tensor, the Ricci scalar and the covariant derivative associated to the
metric $\gamma_{ab}$ respectively. Therefore, the Einstein tensor takes the
form 
\begin{eqnarray}
G_{ij} & = & h_{ij}\left(\tilde{\nabla}_{a}\rho\tilde{\nabla}^{a}\rho+2\rho%
\tilde{\nabla^{2}}\rho-\frac{\rho^{2}}{2}\tilde{R}-1\right)\,,  \notag \\
G_{ia} & = & 0\,,  \label{eq:eten} \\
G_{ab} & = & \tilde{R}_{ab}-\frac{1}{2}\gamma_{ab}\tilde{R}+\frac{3}{\rho^{2}%
}\left[\gamma_{ab}\left(\tilde{\nabla}_{c}\rho\tilde{\nabla}^{c}\rho+\tilde{
\rho\nabla^{2}}\rho-1\right)-\rho\tilde{\nabla}_{b}\tilde{\nabla}_{a}\rho%
\right]\,.  \notag
\end{eqnarray}
The stress-energy tensor (\ref{T1}) for the meron field is given by 
\begin{eqnarray}
T_{ij} & = & \frac{2\lambda^{2}(\lambda-1)^{2}}{e^{2}\rho^{2}}h_{ij}\,, 
\notag \\
T_{ia} & = & 0\,,  \label{eq:tmunu} \\
T_{ab} & =- & \frac{6\lambda^{2}(\lambda-1)^{2}}{e^{2}\rho^{4}}\gamma_{ab}\,,
\notag
\end{eqnarray}
and therefore Einstein equations (\ref{einstein}) yield 
\begin{eqnarray}
\tilde{\nabla}_{a}\rho\tilde{\nabla}^{a}\rho+2\rho\tilde{\nabla}^{2}\rho-%
\frac{\rho^{2}}{2}\tilde{R}+\Lambda\rho^{2}-1 & = & \frac{16\pi G}{
e^{2}\rho^{2}}\lambda^{2}(\lambda-1)^{2} \,,  \label{eq:eeq1} \\
\frac{\rho^{2}}{3}\tilde{G}_{ab}+\gamma_{ab}\left(\tilde{\nabla}_{c}\rho%
\tilde{\nabla}^{c}\rho+\tilde{\rho\nabla^{2}}\rho+\frac{\Lambda\rho^{2}}{3}%
-1\right)-\rho\tilde{\nabla}_{b}\tilde{\nabla}_{a}\rho & = & -\frac{16\pi G}{
e^{2}\rho^{2}}\lambda^{2}(\lambda-1)^{2}\gamma_{ab}\,.  \label{eq:eeq2}
\end{eqnarray}
where we have defined $\tilde{G}{}_{ab}=\tilde{R}_{ab}-\frac{1}{2}\gamma_{ab}%
\tilde{R}$ as the Einstein tensor associated to the metric $\gamma_{ab}$.
Notice that the $(i,j)$ components of the field equations reduce into a
single equation (\ref{eq:eeq1}). 
It should be emphasized that the same reduction of the $(i,j)$ components of
the field equations hold even with the Gauss-Bonnet term on the left hand
side of the field equations, which is given by 
\begin{equation}
H_{\mu \nu} = 2 \big( R R_{\mu \nu} - 2 R_{\mu \rho} R^{\rho}_{\phantom{\rho}
\nu} - 2 R^{\rho \sigma} R_{\mu \rho \nu \sigma} + R_{\mu}^{\phantom{\mu}
\rho \alpha \beta} R_{\nu \rho \alpha \beta} \big) - \frac{1}{2} g_{\mu \nu} %
\big(R^{2} - 4 R_{\alpha \beta} R^{\alpha \beta} + R_{\alpha \beta \gamma
\delta} R^{\alpha \beta \gamma \delta} \big) \ .
\end{equation}
This can be easily shown by observing the following equations 
\begin{eqnarray}
& & R_{ijkm} = \rho^{2} \big( 1 -  \tilde{\nabla}_{a}\rho\tilde{\nabla}^{a}\rho  \big)
 \big( h_{ik}h_{jm} - h_{im} h_{jk} \big) \ , \\
& & R_{ijab} = R_{iabc} = R_{aijk} = 0 \ , \\
& & R_{iajb} = - \rho \ h_{ij} \tilde{\nabla}_{a} \tilde{\nabla}_{b} \rho \ ,
\end{eqnarray}
from which one has $H_{ia} = 0$, and $H_{ij} \propto h_{ij}$. To solve the
Einstein-Yang-Mills equations for meron configurations with the Gauss-Bonnet
term is a valuable task in its own right. In this paper, however, we focus
only on the Einstein-Hilbert action for the gravity sector, and the issues
related to the Gauss-Bonnet gravity will be studied in a separate paper.

\section{Solutions}

\subsection{$D$ = 3}

In three dimensions $\gamma_{ab}=0$ and and $\rho=\rho_{0}$ is constant.
Therefore the metric (\ref{eq:Metric}) is simply given by

\begin{equation}
ds^{2}=\rho_{0}^{2}\sum_{i=1}^{3}\Gamma_{i}\otimes\Gamma_{i}=\frac{
\rho_{0}^{2}}{4}\left(d\tau^{2}+2\cos\theta d\tau
d\varphi+d\theta^{2}+d\varphi^{2}\right)\text{ },
\end{equation}
where we have considered $\psi=\tau$ as the euclidean time. In this case,
Einstein equations (\ref{eq:eeq1}) yield one single algebraic equation for $%
\rho_{0}$, 
\begin{equation}
\Lambda\rho_{0}^{2}-1=\frac{16\pi G}{e^{2}\text{$\rho$}_{0}^{2}}%
\lambda^{2}(\lambda-1)^{2},  \label{eq:eqe1}
\end{equation}
which can be solved for $\Lambda>0$. As Yang-Mills equations (\ref{eq:ymeqs}%
) require $\lambda=\text{1/2}$, equation (\ref{eq:eqe1}) fixes $\rho_{0}$ to
be 
\begin{equation}
\rho_{0}^{2}=\frac{1}{2\Lambda}\left(1\pm\sqrt{1+\frac{4\pi G\Lambda}{e^{2}}}%
\right)\,.  \label{eq:solrho}
\end{equation}
The meronic configuration in this case is defined on the three-sphere with
overall fact $\rho_{0}$ and it is regular and smooth everywhere.

\subsubsection*{Chern-Simons term}

In the three-dimensional case it is possible to find a more general
meron-like solution by adding a Chern-Simons term to the action (\ref{sky2o}%
) and considering 
\begin{equation*}
S_{\mathrm{SU(2)}}=-\frac{1}{8e^{2}}\int d^{D}x\sqrt{g}\mathrm{Tr}%
\left(F^{\mu\nu}F_{\mu\nu}\right)+S_{CS}\,,
\end{equation*}
where the Chern-Simons action for the $SU(2)$ valued gauge field is given by 
\begin{equation}
S_{CS}=\frac{k}{2e^{2}}\int\mathrm{Tr}\left[AdA+\frac{2}{3}A^{3}\right]\ ,
\label{csaction}
\end{equation}
and $k$ is related to the Chern-Simons level\footnote{
There are two possible conventions for the Chern-Simons level $k$: we will 
comment on them in the following sections.}. This modification leads to the
Yang-Mills-Chern-Simons equations 
\begin{equation}
\mathrm{YMCS}^{\mu}=\nabla_{\nu}F^{\mu\nu}+\left[A_{\nu},F^{\mu\nu}\right]%
+k\epsilon^{\nu\rho\sigma}F_{\rho\sigma}=0\,.  \label{YMCS}
\end{equation}
Using \eqref{eq:ymeqs} and \eqref{Fmeron} it is straightforward to check
that 
\begin{equation}  \label{eq:ymcseqs}
\begin{array}{lcl}
\mathrm{YMCS}^{\psi} & = & \dfrac{8\lambda\left(\lambda-1\right)}{\rho^{4} 
\mathrm{sin}\theta}\left(2\lambda-1+k\rho_{0}\right)\left(\mathrm{cos}%
\varphi t_{1}+\mathrm{sin}\varphi t_{2}\right) \\[5pt] 
\mathrm{YMCS}^{\theta} & = & \dfrac{8\lambda\left(\lambda-1\right)\left(2
\lambda-1+k\rho_{0}\right)}{\rho^{4}}\big(-\mathrm{sin}\varphi t_{1}+\mathrm{%
\ cos}\varphi t_{2}\big)\ , \\[5pt] 
\mathrm{YMCS}^{\varphi} & = & -\dfrac{8\lambda(\lambda-1)\left(2\lambda-1+k
\rho_{0}\right)}{\rho^{4}\mathrm{sin}\theta}\big(\mathrm{cos}\theta\mathrm{\
cos}\varphi\ t_{1}+\mathrm{cos}\theta\mathrm{sin}\varphi\ t_{2}+\mathrm{sin}%
\theta\ t_{3}\big)\,,%
\end{array}%
\end{equation}
which leads to 
\begin{equation}
\lambda=\frac{1}{2}(1-k\rho_{0})
\end{equation}
(note that in the Einstein-Yang-Mills case ($k=0$) we get the usual
``meronic'' value $\lambda=1/2$).

Due to its topological nature, the Chern-Simons term does not contribute to
the energy-momentum tensor (\ref{eq:tmunu}). This means that, when (\ref%
{csaction}) is included in the gauge field action, the only modification in
the Einstein equations (\ref{eq:eqe1}) is the value of $\lambda$. In this
case we obtain 
\begin{equation}
\rho_{0}^{2}\Lambda-1=\frac{\pi G}{e^{2}\rho_{0}^{2}}(1-k^{2}\rho_{0}^{2})\
^{2},
\end{equation}
which can be solved for $\Lambda>0$ to give 
\begin{equation}
\rho_{0}^{2}=\frac{e^{2}-2\pi Gk^{2}\pm e\sqrt{e^{2}+4\pi
G\left(\Lambda-k^{2}\right)}}{2e^{2}\Lambda-2\pi Gk^{4}}\ .
\end{equation}
Note that for $\rho_{0}^{2}$ to be positive, one of the following conditions
must hold: 
\begin{eqnarray*}
& & \text{(i)}\ \ e^{2}+4\pi G\left(\Lambda-k^{2}\right)>0\ , \quad2\pi
Gk^{2}>e^{2}\ ,\quad e^{2}\Lambda<\pi Gk^{4}\ , \\
& & \text{(ii)}\ \ e^{2}\Lambda>\pi Gk^{4}\ , \\
& & \text{(iii)}\ \ e^{2}+4\pi G\left(\Lambda-k^{2}\right)=0\ ,
\quad(e^{2}-2\pi Gk^{2})(e^{2}\Lambda-\pi Gk^{4})>0\ , \\
& & \text{(iv)}\ \ e^{2} \Lambda = \pi G k^{4} \ , \quad k^{2}/\Lambda <2 \ .
\end{eqnarray*}

\subsubsection*{Imaginary coupling}

In order for the theory to have a well-defined Lorentzian continuation, the
Euclidean Chern-Simons term must have imaginary coupling ($k\rightarrow ik$, 
$k\in\mathbb{N}$, $i^{2}=-1$). In this case the solutions look very similar
with the difference that the meron parameter $\lambda$ is not real anymore,
: 
\begin{equation*}
\lambda=\frac{1}{2}(1-ikR_{0})\ ,\ \ R_{0}\in\mathbb{R}\ .
\end{equation*}
These configurations represent complex saddle points of the
Einstein-Yang-Mills-Chern-Simons action. In recent years, it has been shown
in many non-trivial examples (see \cite{resurgence1}\ and references
therein; for detailed reviews see \cite{resurgence2,resurgence3,resurgence4}%
) that non-trivial complex saddle points are necessary to give a consistent
non-perturbative definition of the path integral. In particular, when such
complex saddles are not included in the analysis, inconsistencies appear.
Hence, the present results strongly suggest that these gravitating merons
are relevant building blocks to get a consistent path-integral in the
Einstein-Yang-Mills-Chern-Simons case.

\subsubsection*{Euclidean action}

Also in the three-dimensional case the non-perturbative nature of this
configurations is apparent as they depend on $1/e^{2}$. In particular, the
classical Euclidean action $I_{E}$ corresponding to the set of solutions can
be easily computed to give: 
\begin{equation}
I_{E}=h\left(\frac{1}{e^{2}},\Lambda,G,k\right)=\frac{\pi\rho_{0}}{4G}%
\left(\rho_{0}^{2}\Lambda-3\right)+\frac{12\pi^{2}}{e^{2}\rho_{0}}%
\lambda^{2}(\lambda-1)^{2}-\frac{4\pi^{2}k}{e^{2}}\lambda^{2}(\lambda+3)\ .
\end{equation}
The obvious relevance of this result is that, at semi-classical level, the
contribution of this configuration to the path-integral is proportional to $%
Z_{E}$, 
\begin{equation}
Z_{E}\approx\exp\left[-I_{E}\right]\ .
\end{equation}
Therefore, gravitating merons play an important role in the non-perturbative
sector of the theory. This is especially important in the three-dimensional
case in which self-dual instantons do not exist and, consequently, these
Euclidean smooth regular (and with finite actions) configurations can be
quite relevant.

It is also worth to emphasize the remarkable effect of the Chern-Simons term
which supports the existence of gravitating merons with $\lambda\neq1/2$. To
the best of authors knowledge, these are the first examples of smooth merons
with this characteristic. Due to the fact that the Chern-Simons term can
arise upon integrating over Fermionic degrees of freedom, it is natural to
wonder whether one could construct merons with $\lambda\neq1/2$ even with
Fermionic matter fields. We hope to come back on this very interesting
question in a future publication. As it has been already emphasized, in the
case in which the Chern-Simons coupling is taken as $ik$ with $k\in\mathbb{R}
$, the present configurations have to be considered as smooth regular
complex saddle points. Correspondingly, the Euclidean action also gets a
non-trivial imaginary part. These configurations have to be properly
analyzed using resurgence techniques (following \cite%
{resurgence1,resurgence2,resurgence3,resurgence4}). We hope to come back on
this issue in a future publication.

As far as the evaluation of the Euclidean action of the four dimensional
solutions is concerned, it involves some subtleties. The reason is that, in
the presence of a negative cosmological constant, one needs to include
suitable boundary terms to obtain a finite results. The construction of
these boundary terms when topologically non-trivial non-Abelian gauge fields
are present has not been discussed in details in the literature. We hope to
come back on this interesting issue in a future publication.

\subsection{$D$ = 4}

In four dimension we consider only one extra coordinate $z=r$ in (\ref%
{eq:Metric}) and for simplicity we will just take $\gamma_{rr}=1$. The
metric then takes the form 
\begin{equation}
ds^{2}=dr^{2}+\frac{\rho^{2}\left(r\right)}{4}\left(d\tau^{2}+2\cos\theta
d\tau d\varphi+d\theta^{2}+d\varphi^{2}\right).  \label{metrica1}
\end{equation}
where again we have considered $\psi=\tau$ as the euclidean time. Einstein
equations (\ref{eq:eeq1}) and (\ref{eq:eeq2}) are reduced to two ordinary
differential equations 
\begin{eqnarray}
\rho^{\prime2}+2\rho\rho^{\prime\prime}+\Lambda\rho^{2}-1 & = & \frac{\pi G}{
e^{2}\rho^{2}},  \label{eq1} \\
\rho^{\prime2}+\frac{\Lambda}{3}\rho^{2}-1 & = & -\frac{\pi G}{e^{2}\rho^{2}}%
,  \label{eq2}
\end{eqnarray}
where we have already replaced (\ref{eq:solym}). If we plug the equation (%
\ref{eq2}) into (\ref{eq1}), then we have a single ODE of $\rho(r)$, 
\begin{equation}
\rho\rho^{\prime\prime}+\frac{\Lambda}{3}\rho^{2}-\frac{\pi G}{e^{2}\rho^{2}}%
=0.  \label{maineq}
\end{equation}
When the cosmological constant $\Lambda$ is positive, there does not exist
real solution to this equation. Now let us examine the cases of zero and
negative cosmological constants. Similar results have been discussed in \cite%
{maldacena,hosoya,gupta,rey,donets}

\subsubsection*{Case 1: $\Lambda=0$}

When $\Lambda=0$, the solution to (\ref{maineq}) is, 
\begin{equation}
\rho(r)=\frac{1}{e}\sqrt{a(r+b)^{2} + \frac{\pi Ge^{2}}{a}}\ ,
\end{equation}
where $a$ and $b$ are integration constants. This solution satisfies the
equations (\ref{eq1}) and (\ref{eq2}) if 
\begin{equation*}
a=e^{2}.
\end{equation*}
Thus the solution for vanishing cosmological constant is, 
\begin{equation}
\rho(r)=\sqrt{\frac{\pi G}{e^{2}}+(r+b)^{2}}\ .  \label{sol1}
\end{equation}
Hence, these configurations can be interpreted as smooth asymptotically flat
Euclidean wormholes sourced by merons. The size of the throat is
proportional to $1/e^{2}$ thus showing explicitly that the ``opening of the
throat'' is a non-perturbative phenomenon. Moreover, the fact that such
Euclidean wormholes are sourced by Yang-Mills merons (which, by themselves,
represent tunneling between different Gribov vacua \cite{meronsp5}) sheds
considerable light on the physical interpretation of these Euclidean
wormholes. Indeed, the solution is smooth and regular everywhere, the gauge
field is regular and the scale factor $\rho$ is smooth and non-vanishing. In
particular, both asymptotic regions (corresponding to $r\rightarrow\pm\infty$%
) are flat (the wormhole throat being at $r=-b$). Similar Euclidean wormhole
solutions have been studied in \cite{hosoya, gupta, rey, donets}. Examples
of Euclidean wormholes embedded in higher dimensional theories as well as
including the explicit presence of axionic fields have been worked out in 
\cite%
{euclidw1,euclidw2,euclidw3,euclidw4,euclidw5,euclidw6,euclidw7,euclidw8,euclidw9}%
.

\subsubsection*{Case 2: $\Lambda<0$}

When $\Lambda<0$, the solution to (\ref{maineq}) is, 
\begin{equation}
\rho\left(r\right)=\frac{1}{4e}\left[2C_{1}\left(\left(\frac{64\pi Ge^{2}}{
C_{1}^{2}}+C_{2}^{2}\right)\exp\left(2\sqrt{\frac{-\Lambda}{3}}r\right)-%
\frac{3}{4\Lambda}\exp\left(-2\sqrt{\frac{-\Lambda}{3}}r\right)+\sqrt{\frac{%
3 }{-\Lambda}}C_{2}\right)\right]^{1/2},
\end{equation}
where $C_{1}$ and $C_{2}$ are integration constants. The above solution is
real whenever $C_{1}$ is positive. In addition, this solution satisfies the
equations (\ref{eq1}) and (\ref{eq2}) if $C_{1}$ and $C_{2}$ are related by 
\begin{equation*}
C_{1}C_{2}=-4e^{2}\sqrt{\frac{3}{-\Lambda}}.
\end{equation*}
With these conditions, we have the solution $\rho(r)$ given by, 
\begin{equation}
\rho\left(r\right)=\frac{1}{4e}\left[\frac{2}{C_{1}}\left(64\pi Ge^{2}-\frac{
48e^{4}}{\Lambda}\right)\exp\left(2\sqrt{\frac{-\Lambda}{3}}r\right)-\frac{
3C_{1}}{2\Lambda}\exp\left(-2\sqrt{\frac{-\Lambda}{3}}r\right)+\frac{24e^{2}%
}{\Lambda}\right]^{1/2}\,.
\end{equation}
Let us notice that the argument of the square root is positive definite, and
its minimum value 
\begin{equation}
\rho_{min}=\frac{\sqrt{6}}{2\sqrt{-\Lambda}}\left[\sqrt{1-\frac{4\pi
G\Lambda }{3e^{2}}}-1\right]^{1/2}  \label{min}
\end{equation}
occurs at 
\begin{equation}
r=\frac{1}{2}\sqrt{\frac{3}{-\Lambda}}\mathrm{Log}\left(\frac{\sqrt{3}C_{1}}{
8e\sqrt{3e^{2}-4\pi G\Lambda}}\right),  \label{minat}
\end{equation}
if the right hand side of (\ref{minat}) is positive. Therefore, if we choose
a sufficiently small positive constant $C_{1}$, then the corresponding
solution is regular and smooth everywhere for any $r\in\mathbb{R}$. \newline
In these cases both asymptotic regions (namely, $r\rightarrow\pm\infty$) are
(the Euclidean version of) AdS.

Thus, both in \textit{Case 1} and in \textit{Case 2} described above the
gravitating merons can be interpreted as smooth Euclidean wormholes
interpolating between the vacua of the theory. It is also worth to emphasize
that also in this case the (size of the) wormhole throat is non-perturbative
in the Yang-Mills coupling $e^{2}$ (as it depends on $1/e^{2}$: see Eqs. (%
\ref{sol1}) and (\ref{min})). Consequently, the present configurations will
be relevant in the non-perturbative sector of Einstein-Yang-Mills theory.

\subsection{$D$ = 5}

Solutions with constant $\rho$ analogous to the three-dimensional one (\ref%
{eq:solrho}) previously constructed cannot be generalized to four
dimensions, as in that case the equations (\ref{eq:eeq1}) and (\ref{eq:eeq2}%
) do not admit solutions for constant $\rho$. For $D>4$, however, the
warping factor $\rho$ can be taken as a constant $\rho_0$. In five
dimensions we can consider coordinates $z^{a}=(\tau,r)$, were $\text{$\tau$}$
is the Euclidean time and $r$ a radial coordinate. The simplest solutions of
the form (\ref{eq:Metric}) can be obtained by considering a two-dimensional
metric $\gamma_{ab}$ with constant curvature $\tilde{R}=K$ and 
\begin{equation*}
\gamma_{ab}=\left(%
\begin{array}{cc}
r^{2} & 0 \\ 
0 & \frac{1}{1-\frac{K}{2}r^{2}}%
\end{array}%
\right)\,.
\end{equation*}
In that case, Einstein equations (\ref{eq:eeq1}) and (\ref{eq:eeq2}) take
the form 
\begin{gather}
\left(\frac{K}{2}-\Lambda\right)\rho_{0}^{2}+1+\frac{\text{$\pi$}G}{
e^{2}\rho_{0}^{2}}=0\,,  \label{eq:eeq5d1} \\
\frac{\Lambda\rho_{0}^{2}}{3}-1+\frac{\text{$\pi$}G}{e^{2}\rho_{0}^{2}}=0\,.
\label{eq:eeq5d2}
\end{gather}
Eq. \eqref{eq:eeq5d1} fixes $K$ in terms of $\rho_{0}$, $\Lambda$, $G$ and $%
e $, 
\begin{equation*}
K=2\Lambda-\frac{2}{\rho_{0}^{2}}\left(1+\frac{\text{$\pi$}G}{
e^{2}\rho_{0}^{2}}\right)\,,
\end{equation*}
while Eq. \eqref{eq:eeq5d2} determines $\rho_{0}^{2}$:

\begin{itemize}
\item For $\Lambda>0$ and $\frac{\text{$4\pi$}G\Lambda}{3e^{2}}\leq1$,
\end{itemize}

\begin{equation}
\rho_{0}^{2}=\frac{3}{2\Lambda}\left[1\pm\sqrt{1-\frac{\text{$4\pi$}G\Lambda 
}{3e^{2}}}\right]\,.  \label{eq:rhoconstant5d1}
\end{equation}

\begin{itemize}
\item For $\Lambda=0$ ,
\end{itemize}

\begin{equation}
\rho_{0}^{2}=\frac{\text{$\pi$}G}{e^{2}}\,.  \label{eq:rhoconstant5d2}
\end{equation}

\begin{itemize}
\item For $\Lambda<0$,
\end{itemize}

\begin{equation}
\rho_{0}^{2}=\frac{3}{2\Lambda}\left[1-\sqrt{1-\frac{\text{$4\pi$}G\Lambda}{
3e^{2}}}\right]\,.  \label{eq:rhoconstant5d3}
\end{equation}
As in the three-dimensional case, one could be tempted to add a
five-dimensional Chern-Simons term to the Yang-Mills actions (\ref{sky2o}).
However the five-dimensional Chern-Simons equations are proportional to $%
\epsilon^{\mu\nu\rho\sigma\lambda}F_{\nu\rho}F_{\sigma\lambda}$ which
vanishes for the the meron field-strength (\ref{Fmeron}). The same argument
holds in higher odd-dimensional cases.

\subsection{Higher dimensions}

Solutions of the form (\ref{eq:rhoconstant5d1}) can be easily extended to
arbitrarily higher dimensions. In fact, for $\rho=\rho_{0}$ a constant, and $%
\gamma_{ab}$ a $d$-dimensional metric. Einstein equations (\ref{eq:eeq1})
and (\ref{eq:eeq2}) reduce in general to 
\begin{gather}
\left(\frac{\tilde{R}}{2}-\Lambda\right)\rho_{0}^{2}+1+\frac{\text{$\pi$}G}{
e^{2}\rho_{0}^{2}}=0 \,,  \label{eq:eeqanyd1} \\
\tilde{G}_{ab}+\left[\Lambda+\frac{3}{\rho_{0}^{2}}\left(\frac{\text{$\pi$}G%
}{e^{2}\rho_{0}^{2}}-1\right)\right]\gamma_{ab}=0\,.  \label{eq:eeqanyd2}
\end{gather}
The first equation implies that the Ricci tensor $\tilde{R}$ for the metric $%
\gamma_{ab}$ is constant, while the second equation can be written as the
Einstein equations for $\gamma_{ab}$ with an effective cosmological
constant: 
\begin{equation*}
\tilde{\Lambda}=\Lambda+\frac{3}{\rho_{0}^{2}}\left(\frac{\text{$\pi$}G}{
e^{2}\rho_{0}^{2}}-1\right)\,.
\end{equation*}
This means that in any dimension $D=d+3$ with $d>2$, the metric $\gamma_{ab}$
is an Einstein manifold with cosmological constant $\tilde{\Lambda}$, i.e, 
\begin{equation*}
\tilde{R}_{ab}=\frac{2\tilde{\Lambda}}{d-2}\gamma_{ab}\,,
\end{equation*}
which means that the Ricci scalar $\tilde{R}$ is given by 
\begin{equation*}
\tilde{R}=\frac{2d}{d-2}\left[\Lambda+\frac{3}{\rho_{0}^{2}}\left(\frac{ 
\text{$\pi$}G}{e^{2}\rho_{0}^{2}}-1\right)\right]\,.
\end{equation*}
Plugging this back in Eq. (\ref{eq:eeqanyd1}) we find $\rho_{0}^{2}$ to be:

\begin{itemize}
\item For $\Lambda>0$ and $\frac{\text{$4\pi$}G\Lambda(2d-1)}{e^{2}(d+1)^{2}}
\leq1$,
\end{itemize}

\begin{equation}
\rho_{0}^{2}=\frac{d+1}{2\Lambda}\left[1\pm\sqrt{1-\frac{\text{$4\pi$}
G\Lambda(2d-1)}{e^{2}(d+1)^{2}}}\right]\,.  \label{eq:rhoconstantanyd1}
\end{equation}

\begin{itemize}
\item For $\Lambda=0$ ,
\end{itemize}

\begin{equation}
\rho_{0}^{2}=\frac{\text{$\pi$}G(2d-1)}{e^{2}(d+1)}\,.
\label{eq:rhoconstantanyd2}
\end{equation}

\begin{itemize}
\item For $\Lambda<0$,
\end{itemize}

\begin{equation}
\rho_{0}^{2}=\frac{d+1}{2\Lambda}\left[1\pm\sqrt{1-\frac{\text{$4\pi$}
G\Lambda(2d-1)}{e^{2}(d+1)^{2}}}\right]\,.  \label{eq:rhoconstantanyd3}
\end{equation}
The fact that any $d$-dimensional Einstein manifold with cosmological
constant $\tilde{\Lambda}$ and constant provides a solution for $\gamma_{ab}$
is very interesting. In higher dimensions one could use different known
solutions plus the three-sphere to construct Euclidean geometries supporting
meron-like configurations of the form \eqref{Ameron}. One interesting
example would be, for instance, to use the Euclidean Schwarzschild-AdS or
Euclidean Kerr-AdS black holes in four dimensions as the metric $\gamma_{ab}$%
, to form a seven-dimensional black brane with three compact dimensions. 
It would be also very interesting to construct solutions with a non-constant
and regular warp factor. This task, however, is quite non-trivial and it is
likely that some extra ingredients are required to achieve it. We hope to
come back on this issue in a future publication.

\section{Conclusions}

Analytic smooth configurations of Euclidean Einstein-Yang-Mills system have
been constructed. The ansatz for the gauge field is of meron-type: it is
proportional to a pure gauge (with a suitable parameter $\lambda$ which is
determined by solving the field equations). The smooth gauge transformation
used to construct the meron cannot be deformed continuously to the identity
as it possesses a non-vanishing winding number. In the three dimensional
case, the solution is smooth and the spatial geometry is a three-sphere. The
effects of the inclusion of a Chern-Simons term can be studied explicitly.
Interestingly enough, one of the effects of the Chern-Simons term is that,
unlike what happens in the pure Yang-Mills case, the parameter $\lambda$ is
in general different from $1/2$: the value of $\lambda$ in the 3D
Yang-Mills-Chern-Simons case depends explicitly on the Chern-Simons
coupling. In dimensions greater than three, one gets $\lambda=1/2$. In four
dimensions the corresponding geometry can be interpreted as a smooth
Euclidean wormhole interpolating between different vacua of the theory
(thus, extending the usual flat interpretation of merons). In five
dimensions regular meron-like configurations have been found, where the
metric is given by the three-sphere times a constant curvature space. This
last result can be extended to arbitrary higher dimensions where the metric
is given by the warped product the three-sphere with any solution of the $%
(D-3)$-dimensional Einstein equations in vacuum with an effective
cosmological constant. In all theses cases, the coupling of the meron with
general relativity ``regularizes" the configurations. Namely, Yang-Mills
configurations (which on flat spaces would be singular) become regular when
the coupling with general relativity is considered. This remarkable effect
could be named gravitational catalysis of merons. One of the consequences of
this fact is that, while in the flat case the Euclidean action of merons is
divergent (so that a single meron gives vanishing contribution to the
semi-classical path integral), gravitating merons can be smooth and regular
and, consequently, they can give a non-vanishing contribution to the
semi-classical path integral (as the present examples clearly show). In
Cho's approach we can express the vacuum potential $\Omega_{\mu}=U^{-1}%
\partial_{\mu}U$ explicitly with $\hat{n}=(n_{1},n_{2},n_{3})$, and express
the ansatz (11) solely by $\hat{n}$. With this we can obtain the same result
using $\hat{n}$ \cite{cho1,cho2,cho3,cho4,cho5}. A very interesting issue
(on which we hope to come back in a future publication) is the resurgence
analysis (along the lines of \cite{resurgence1}) of the complex regular
meron-like saddle points which appear in the
Einstein-Yang-Mills-Chern-Simons case when the Chern-Simons coupling
constant is taken as $ik$.

\subsection*{Acknowledgements}

We warmly thank Y. M. Cho, A. Gomberoff, M. Riquelme, J. H. Yoon and J.
Zanelli for illuminating discussions and important suggestions. We also
would like to thank the anonimous referee for her/his valuable comments.
This work has been funded by the Fondecyt grants 1160137, 3160581 and by the
National Research Foundation of Korea funded by the Ministry of Education of
Korea (Grants 2015-R1D1A1A01-057578, 2015-R1D1A1A01-059407). The Centro de
Estudios Científicos (CECs) is funded by the Chilean Government through the
Centers of Excellence Base Financing Program of Conicyt.


\begin{thebibliography}{99}
\bibitem{manton} {\small N.~Manton and P.~Sutcliffe, }\textit{{\small %
{}{}Topological Solitons}}{\small {}{}, Cambridge University Press, 
Cambridge, (2007). }

\bibitem{confinement} {\small J. Greensite, }\textit{{\small {}{}An
Introduction to the Confinement Problem}}{\small {}{}, Lecture Notes in 
Physics, Springer (2011). }

\bibitem{meronp1} {\small V. de Alfaro, S. Fubini and G. Furlan, Phys. Lett.
\textbf{B 65}, 163 (1976). }

\bibitem{meronp2} {\small C. G. Callan, Jr., R. F. Dashen and D. J. Gross, 
Phys. Rev. \textbf{D 19}, 1826 (1979); Phys. Rev. \textbf{D 17}, 2717
(1978); Phys. Lett. \textbf{B 66}, 375 (1977). }

\bibitem{meronp3} {\small J. Glimm and A. M. Jaffe, Phys. Rev. \textbf{Lett.
40}, 277  (1978). }

\bibitem{meronp4} {\small F. Lenz, J. W. Negele and M. Thies, Phys. Rev. 
\textbf{D 69},  074009 (2004) {[}hep-th/0306105{]}. }

\bibitem{meronsp5} {\small A. Actor, Rev. Mod. Phys. \textbf{51}, 461
(1979). }

\bibitem{rubakov} {\small V. A. Rubakov, M. E. Shaposhnikov, }\textit{\ 
{\small {}{}Electroweak Baryon Number Non-Conservation in the Early Universe
and in High Energy Collisions}}{\small {}{}, }\textit{{\small {}{}Phys. Usp.}
}{\small {}{} }\textbf{{\small {}{}39}}{\small {}{}, 461-502, (1996); arXiv:
hep-ph/9603208v2. }

\bibitem{teitelboim} {\small P. Cordero, C. Teitelboim Ann. Phys. \textbf{100%
},  607-631 (1976). }

\bibitem{mbh} {\small F. Canfora, F. Correa, A. Giacomini, J. Oliva,  Phys.
Lett. \textbf{B 722}, 364-371 (2013). }

\bibitem{tmym1} {\small S. Deser, R. Jackiw, S. Templeton, }\textit{{\small %
{}{}Phys. Rev. Lett.}}{\small {}{} }\textbf{{\small {}{}48}}, {975  (1982). }

\bibitem{tmym2} {\small S. Deser, R. Jackiw, S. Templeton, }\textit{{\small %
{}{}Ann. Physics}}{\small {}{} }\textbf{{\small {}{}140}},{\small {}{}  372
(1982); S. Deser, R. Jackiw, S. Templeton, }\textit{{\small {}{}Ann. Physics}%
} {\small {}{} }\textbf{{\small {}{}185}}{\small {}{}, 406 (1988); }\textit{%
{\small {}{}Ann. Physics}}, \textbf{\ {\small {}{}281}}{\small {}{}, 409
(2000). }

\bibitem{CSF1} {\small S. Deser, L. Griguolo, D. Seminara, }\textit{{\small %
{}{}Comm. Math. Phys}}{\small {}{}. }\textbf{{\small {}{}197}},{\small {}{}
433--450  (1998). }

\bibitem{CSF2} {\small S. Deser, L. Griguolo, D. Seminara, }\textit{{\small %
{}{}Phys. Rev.}}{\small {}{} }\textbf{{\small {}{}D 57}},{\small {}{}
7444--7459 (1998). }

\bibitem{reviewCS} {\small G. Dunne, }\textit{{\small {}{}Les Houches
Lectures}}{\small {}{} (1998); arXiv: 9902.115. }

\bibitem{ourann} {\small F. Canfora, A. Gomez, S. P. Sorella, D. 
Vercauteren, }\textit{{\small {}{}Ann. Physics}}{\small {}{} }\textbf{\ 
{\small {}{}345}}, {\small {}{}166 (2014). }

\bibitem{numym1} {\small P. Bizon, Phys. Rev. \textbf{Lett. 64}, 2844
(1990). }

\bibitem{numym2} {\small P. Breitenlohner, P. Forgacs and D. Maison, Nucl. 
Phys. \textbf{B 383}, 357 (1992); Nucl. Phys. \textbf{B 442}, 126 (1995). }

\bibitem{numym3} {\small K. -M. Lee, V. P. Nair and E. J. Weinberg, Phys. 
Rev. \textbf{Lett. 68}, 1100 (1992); Phys. Rev. \textbf{D 45}, 2751 (1992);
Gen. Relativ.  Gravit. \textbf{24}, 1203 (1992). }

\bibitem{numym4} {\small P. C. Aichelburg and P. Bizon, Phys. Rev. \textbf{D
48}, 607  (1993). }

\bibitem{numym5} {\small T. Tachizawa, K. -I. Maeda and T. Torii, Phys. Rev.
\textbf{D 51}, 4054 (1995). }

\bibitem{canfora} {\small F. Canfora, P.~Salgado-Rebolledo, }\textit{{\small %
{}{}Phys. Rev.}}{\small {}{} }\textbf{{\small {}{}D 87}}{\small {}{}, 045023
(2013). }

\bibitem{canfora2} {\small F. Canfora, H.~Maeda, }\textit{{\small {}{}Phys.
Rev.}}{\small {}{} }\textbf{{\small {}{}D 87}}{\small {}{}, 084049 (2013). }

\bibitem{canfora3} {\small F. Canfora, }\textit{{\small {}{}Phys. Rev.}} 
{\small {}{} }\textbf{{\small {}{}D 88}}{\small {}{}, 065028 (2013). }

\bibitem{canfora4} {\small F. Canfora, F. Correa, J. Zanelli, }\textit{\ 
{\small {}{}Phys. Rev.}}{\small {}{} }\textbf{{\small {}{}D 90}}{\small %
{}{}, 085002 (2014). }

\bibitem{canfora4bis} {\small F. Canfora, A. Giacomini, S. Pavluchenko,}%
\textit{\ {\small {}{}Phys. Rev.}}{\small {}{} }\textbf{{\small {}{}D 90}}%
{\small , 043516 (2014).}

\bibitem{yang1} {\small S. Chen, Y. Li, Y. Yang, Phys. Rev. \textbf{D 89}, 
025007 (2014). }

\bibitem{canfora5} {\small F. Canfora, M. Kurkov, M. Di Mauro, A. Naddeo, } 
\textit{{\small {}{}Eur.Phys.J.}}{\small {}{} }\textbf{{\small {}{}C75}} 
{\small {}{}, 443 (2015). }

\bibitem{yang2} {\small S. Chen, Y. Yang, Nucl. Phys. \textbf{B 904}, 470
(2016). }

\bibitem{canfora6} {\small E. Ayon-Beato, F. Canfora, J. Zanelli, Phys. 
Lett. \textbf{B 752}, 201-205 (2016). }

\bibitem{canfora7} {\small F. Canfora, G. Tallarita, Phys. Rev. \textbf{D 94}%
,  025037 (2016).}

\bibitem{canfora8} {\small F. Canfora, G. Tallarita, Phys. Rev. \textbf{D 91}%
,  085033 (2015). }

\bibitem{canfora9} {\small F. Canfora, G. Tallarita, JHEP \textbf{1409}, 136
(2014). }

\bibitem{canfora10} {\small F. Canfora, G. Tallarita, Nucl. Phys. \textbf{B
921},  394 (2017).}

\bibitem{canfora11} {\small F. Canfora, A. Paliathanasis, T. Taves, J. 
Zanelli, Phys. Rev. \textbf{D 95}, 065032 (2017).}

\bibitem{canfora12} {\small F. Canfora, N. Dimakis, A. Paliathanasis, Phys. 
Rev. \textbf{D 96}, 025021 (2017).}

\bibitem{canfora13} {\small P.D. Alvarez, F. Canfora, N. Dimakis, A.
Paliathanasis, Phys. Lett. \textbf{B 773}, 401-407 (2017).}

\bibitem{cho1} {\small Y. M. Cho, Phys. Rev. \textbf{Lett. 46}, 302 (1981);
Phys.  Rev. \textbf{Lett. 87}, 252001 (2001); Phys. Lett. \textbf{B 603}, 88
(2004);  Phys. Lett. \textbf{B 644}, 208 (2007). }

\bibitem{cho2} {\small Y.M. Cho, D. Maison, Phys. Lett. \textbf{B 391},
360-365 (1997). }

\bibitem{cho3} {\small K. Kimm, J. H. Yoon, Y. M. Cho, Eur. Phys. J. \textbf{%
C 75}, 67 (2015). }

\bibitem{cho4} {\small Y. M. Cho, Franklin H. Cho, and J. H. Yoon, Phys. 
Rev. \textbf{D 87}, 085025 (2013). }

\bibitem{cho5} {\small Y.M. Cho, Kyoungtae Kimm, J. H. Yoon, Pengming Zhang,
Euro Phys. J.} \textbf{{\small {}C77}}{\small {}, 88 (2017). }

\bibitem{hosoya} {\small A. Hosoya and W. Ogura, Phys. Lett. \textbf{B 225},
117-120 (1989). }

\bibitem{gupta} {\small A. K. Gupta, J. Hughes, J. Preskill and M. B. Wise, 
Nucl. Phys. \textbf{B 333}, 195 (1990).}

\bibitem{rey} {\small S. Rey, Nucl. Phys. \textbf{B 336}, 146 (1990).}

\bibitem{donets} {\small E. E. Donets and D.V. Gal'tsov, Phys.Lett.  \textbf{%
B 296}, 311-315 (1992).}

\bibitem{maldacena} {\small J. Maldacena and L. Maoz, JHEP \textbf{0402},
053 (2004).}

\bibitem{euclidw1} {\small S. B. Giddings and A. Strominger, Nucl. Phys. 
\textbf{B 306}, 890 (1988).}

\bibitem{euclidw2} {\small G. V. Lavrelashvili, V. A. Rubakov and P. G. 
Tinyakov, JETP Lett. \textbf{46}, 167 (1987) {[}Pisma Zh. Eksp. Teor. Fiz. 
\textbf{46}, 134  (1987){]}.}

\bibitem{euclidw3} {\small S. W. Hawking, Phys. Rev. \textbf{D 37}, 904
(1988).}

\bibitem{euclidw4} {\small N. Arkani-Hamed, J. Orgera, J. Polchinski, JHEP  
\textbf{0712}, 018 (2007).}

\bibitem{euclidw5} {\small E. Bergshoeff, A. Collinucci, A. Ploegh, S. 
Vandoren, T. Van Riet, JHEP \textbf{0601}, 061 (2006).}

\bibitem{euclidw6} {\small A. Bergman and J. Distler, arXiv:0707.3168 {[}
hep-th{]}.}

\bibitem{euclidw7} {\small S. R. Coleman, Nucl. Phys. \textbf{B 307}, 867
(1988).}

\bibitem{euclidw8} {\small S. B. Giddings and A. Strominger, Nucl. Phys. 
\textbf{B 307}, 854 (1988).}

\bibitem{euclidw9} {\small T. Hertog, M. Trigiante, T. Van Riet, 
arXiv:1702.04622v1 {[}hep-th{]}.}

\bibitem{resurgence1} {\small G. V. Dunne, M. Unsal,\ JHEP \textbf{12} 002
(2016);  arXiv:1609.05770 {[}hep-th{]}.}

\bibitem{resurgence2} {\small T. Sulejmanpasic, 
https://math.tecnico.ulisboa.pt/seminars/download.php?fid=125}

\bibitem{resurgence3} {\small G. V. Dunne, M. Unsal, }\textit{{\small {}What
is QFT? Resurgent trans-series, Lefschetz thimbles, and new exact saddles}} 
{\small {}, arXiv:1511.05977v1.}

\bibitem{resurgence4} {\small D. Dorigoni, }\textit{{\small {}An
Introduction to Resurgence, Trans-Series and Alien Calculus}}{\small {}, 
arXiv:1411.3585v2. }
\end{thebibliography}
\end{document}